
\input phyzzx.tex

%
\tolerance=500000 \overfullrule=0pt

        \def\cmp{Commun. Math. Phys.}
     
    \def\jgp{J. Geom. Phys.} 
\def\jmp{J. Math. Phys.}

\def\mrl{Math. Res. Lett.}  \def\np{Nucl. Phys.}
\def\pl{Phys. Lett.}

\def\mani{\cal{M}}  
\def\ad{\hbox{\rm ad}}

\def\ex{{\hbox{\rm e}}}  
\def\tr{{\hbox{\rm Tr}}}
\def\too{\longrightarrow}
\def\half{{1\over 2}} 

\def\to{\rightarrow}

\def\sqr#1#2{{\vcenter{\vbox{\hrule height.#2pt
        \hbox{\vrule width.#2pt height#1pt \kern#1pt
           \vrule width.#2pt}
        \hrule height.#2pt}}}}

\def\deriv{{\cal D}}
%
%

\def\raiz{\sqrt{2}}

\def\dalpha{{\dot\alpha}}

%
%

%
%
\font\upright=cmu10 
\font\sans=cmss12
\def\ssf{\sans}

\def\ZZ{\hbox{\rlap{\ssf Z}\kern 2.7pt {\ssf Z}}}
\def\mt{\rlap{\ssf T}\kern 3.0pt{\ssf T}}
\def\identity{{\upright\rlap{1}\kern 2.0pt 1}}
\def\inbar{\vrule height1.5ex width.4pt depth0pt}
\def\CC{\relax\,\hbox{$\inbar\kern-.3em{\rm C}$}}
\def\QQ{\relax\,\hbox{$\inbar\kern-.3em{\rm Q}$}}
\def\RR{\relax{\rm I\kern-.18em R}}
\def\PP{\relax{\rm I\kern-.18em P}}
\def\CP{\CC\PP}

\pubnum={CERN-TH/97-316 \cr US-FT-33/97 \cr hep-th/9711132}
\date={November, 1997}
\pubtype={}
\titlepage

\title{MASS PERTURBATIONS IN TWISTED $N=4$ SUPERSYMMETRIC GAUGE THEORIES} \author{J. M. F. Labastida$^{a,b}$ {\twelverm and}  
Carlos Lozano$^{b}$
\foot{e-mail: lozano@fpaxp1.usc.es}}
\address{$^{a}$ Theory Division, CERN\break CH-1211 Geneva 23,
Switzerland\break\break $^{b}$ Departamento de F\'\i sica de Part\'\i culas\break Universidade de Santiago de Compostela\break E-15706 Santiago de Compostela, Spain}  

\abstract{Mass perturbations of the twisted $N=4$ supersymmetric gauge theory considered by Vafa and Witten to test $S$-duality are studied for the case of K\"ahler four-manifolds. It is shown that the resulting mass-perturbed theory can be regarded as an equivariant extension associated to a $U(1)$ symmetry of the twisted theory, which is only present for K\"ahler manifolds. In addition, it is shown that the partition function, the only topological invariant of the theory,  remains invariant under the perturbation.

\vskip1cm
\noindent CERN-TH/97-316
\vskip-0.5cm
\noindent November 1997}

\endpage  
\pagenumber=1


\chapter{Introduction}

Topological quantum field theory \REF\tqft{E. Witten, ``Topological Quantum
Field Theory"\journal\cmp&117 (88)353.} [\tqft] has become a very useful 
framework to
make predictions in differential topology, and to test some of the recent ideas
emerged in the context of duality as a symmetry of field theories with extended
supersymmetry
\REF\swequ{N. Seiberg and E. Witten, ``Electric-Magnetic Duality, Monopole 
Condensation, and Confinement in $N=2$ Supersymmetric Yang-Mills Theory", 
{\sl Nucl. Phys.} {\bf B426} (1994), 19,  
Erratum-{\sl ibid}.{\bf B430} (1994), 485; hep-th/9407087.}
\REF\swotro{N. Seiberg and E. Witten, ``Monopoles, Duality and Chiral Symmetry
Breaking in N=2 Supersymmetric QCD", {\sl Nucl. Phys.}
{\bf B431}  (1994), 484; hep-th/9408099.} [\swequ, \swotro]. 
The most celebrated examples are the prediction made by Witten 
\REF\monop{E. Witten, ``Monopoles and Four-Manifolds"\journal\mrl&1 (94)769; hep-th/9411102.} [\monop] stating that the Donaldson invariants of 
four-manifolds can be expressed in terms of the Seiberg-Witten invariants, 
and the strong-coupling test of $S$-duality carried out by Vafa and Witten 
\REF\vafa{C. Vafa and E. Witten, ``A Strong Coupling Test of S-Duality"\journal\np&B431 (94)3; hep-th/9408074.} [\vafa] 
making use of a twisted four-dimensional $N=4$ supersymmetric gauge theory. 

The topological quantum field theory leading to the Donaldson invariants 
can be regarded as a twisted version of the $N=2$ supersymmetric pure gauge 
theory. This theory, now known as the Donaldson-Witten theory, possesses 
observables whose correlation functions correspond to those invariants. These 
quantities are independent of the coupling constant of the theory and can thus be studied in both the weak and the strong coupling limits.  By going to 
the weak coupling limit, it can be shown that these correlation functions do 
in fact correspond to the Donaldson polynomials of four-manifolds. These are 
basically intersection numbers on classical instanton moduli spaces, which are sensitive to the differentiable structure of the four-manifold. However, while the weak-coupling analysis provides an astonishing
link to the Donaldson theory, it is not possible to perform explicit 
calculations without using the standard methods inherent in the Donaldson 
theory. A natural way around is to exploit the coupling constant independence 
of the theory to study it in the strong-coupling limit. However, this 
analysis requires a precise knowledge of the infrared behaviour of the
$N=2$ supersymmetric gauge theory, and this was out of reach until the 
explicit solution of Seiberg
and Witten [\swequ, \swotro]. The understanding of the 
strong-coupling dynamics of the $N=2$ supersymmetric gauge theory triggered 
a major breakthrough, by turning the problem
of calculating correlation functions in a twisted supersymmetric gauge theory 
into one of counting solutions of Witten's Abelian monopole equations 
[\monop]. This approach makes possible an explicit calculation of the 
Donaldson polynomials in terms of Seiberg-Witten invariants. 
A similar structure has been proposed for a generalization of the 
Donaldson-Witten theory known as the non-Abelian monopole theory 
\REF\marmon{J. M. F. Labastida and  M. Mari\~no, ``Non-Abelian Monopoles on 
Four-Manifolds"\journal\np&B448 (95)373; hep-th/9504010.}
\REF\corea{S. Hyun, J. Park, J.-S. Park, ``Topological QCD"\journal\np&B453 (95)199; hep-th/9503020.} 
[\marmon,\corea]. Recently, these results have been reviewed in 
\REF\laplata{J. M. F. Labastida and C. Lozano, ``Lectures on Topological 
Quantum Field Theory", hep-th/9709192.} [\laplata], and they have been extended 
and rederived in a more general framework in 
\REF\wimoore{G. Moore and E. Witten, ``Integration over the $u$-plane in 
Donaldson Theory", hep-th/9709193.} [\wimoore].

There is, however, a complementary approach due to Witten 
\REF\wijmp{E. Witten, ``Supersymmetric Yang-Mills Theory on a Four-Manifold"\journal\jmp&35 (94)5101; hep-th/9403195.} [\wijmp]  
(sometimes referred to as the ``abstract" approach, as opposite to the 
``concrete" approach described in the previous paragraph), which works only 
on K\"ahler manifolds and relies heavily on standard results on $N=1$ 
supersymmetric gauge theories, such as gluino condensation and chiral 
symmetry breaking. But
this is doubly as good, for the agreement found between 
the proposed formulas in the topological field theory 
and previously known mathematical results gives support 
to the conjectured picture in the physical theory. The same idea has
subsequently been applied to other $N=2$ supersymmetric gauge theories, as 
in \REF\marpol{J. M. F. Labastida and M.
Mari\~no, ``Polynomial Invariants for $SU(2)$ Monopoles"\journal\np&B456 (95)633; hep-th/9507140.} [\marpol], to obtain
explicit results for the topological invariants associated 
to non-Abelian monopole theory, and also to one of the twisted $N=4$ 
supersymmetric gauge theories [\vafa], to make an explicit computation of 
the partition function of the theory on K\"ahler 
manifolds. 

The way the construction works is the following. When formulated on 
K\"ahler manifolds, the number of BRST charges of a topological quantum 
field theory is doubled, in such a way that, for example,  the 
Donaldson-Witten theory has an enhanced $N_T=2$ topological symmetry  
on K\"ahler manifolds, while the Vafa-Witten theory has $N_T=4$ topological 
symmetry. In either
case, one of the BRST charges comes from the underlying $N=1$ 
subalgebra which corresponds to the formulation of the physical theory in 
$N=1$ superspace. By suitably adding mass terms for some of the chiral 
superfields in the theory, one can break the extended ($N=2$ or $N=4$) 
supersymmetry of the physical theory down to $N=1$. For the reason sketched above, the corresponding
twisted massive theory on K\"ahler manifolds should still retain at least one topological symmetry. One now exploits the metric independence of the topological theory. By scaling up the metric in the topological theory, $g_{\mu\nu}\to tg_{\mu\nu}$, one can take the limit $t\to\infty$. In this limit, the metric on $X$ becomes nearly flat. As the twisted and the physical
theories coincide on flat and hyper-K\"ahler manifolds,
this means that in the $t\to\infty$ limit the predictions of the perturbed topological theory should coincide with those of the physical (massive) theory. But the $t\to\infty$ limit also corresponds to the infrared limit of the physical $N=1$ supersymmetric gauge theory, in which the massive superfields can be integrated out, so one is left with an effective massless $N=1$ supersymmetric gauge theory (possibly) coupled to $N=1$ supersymmetric matter, whose infrared behaviour is --hopefully--
easier to deal with. In this way, the computations in the topological field theory can be reduced to the analysis of contributions from the  vacua of the associated $N=1$ supersymmetric gauge theory.

There is, however, an obvious drawback to this construction. The introduction
of a mass perturbation may (and in general will) distort the original
topological field theory. This poses no problem in the case of the Donaldson-Witten
theory, as Witten was able to prove that the perturbation is topologically
trivial, in the sense that it affects the theory in an important but
controllable way [\wijmp]. 
\REF\coreados{J.-S. Park, ``Monads and D-Instantons"\journal\np&B493 (97)198; hep-th/9612096.} 
However, the arguments presented there do not 
carry over to other, more general situations, so one has to repeat the 
analysis case by case. In the case of the Vafa-Witten theory, the required 
perturbation 
gives rise to an a priori different theory, in fact an equivariant extension 
of the original theory with respect to a $U(1)$ action on the moduli space,  
which is present only on K\"ahler manifolds\foot{This has been pointed out by J.-S. Park in [\coreados].}. We do not know whether the 
theories are actually different or not. But in any case, we are primarily 
interested in calculating the partition function of the theory which, as we 
will argue below, is actually invariant under the perturbation. 

The main purpose of this paper is to show that, as assumed in [\vafa], the abstract approach can be applied successfully to the Vafa-Witten theory on K\"ahler manifolds. In the process we find that the mass-perturbed theory involved in this approach can be regarded as an equivariant extension associated to a certain $U(1)$ symmetry.

The paper is organized as follows. In sect. 2 we review the twisting procedure involved in $N=4$ supersymmetric gauge theories, and formulate the Vafa-Witten theory on K\"ahler manifolds. In sect. 3 we analyse the possible mass perturbations of the theory, and show that the partition function associated to the mass-perturbed Vafa-Witten theory remains invariant. In sec. 4 we reformulate the mass-perturbed theory as an equivariant extension associated to a $U(1)$ symmetry present in the Vafa-Witten theory on K\"ahler manifolds.
Finally, in sect. 5 we present our conclusions.

\endpage


\chapter{Twisting of $N=4$ supersymmetric gauge theory on K\"ahler manifolds}

In this section we  review  some aspects of the twisting of four-dimensional $N=4$ supersymmetric gauge  theories, and we present the form of one of the twisted theories, the Vafa-Witten theory, for the case of K\"ahler manifolds.

\section{$N=4$ supersymmetric gauge theory}

We begin by recalling several generalities about the $N=4$ supersymmetric 
gauge theory on flat ${\RR}^4$.  From the point of view of $N=1$ superspace, the theory contains
 one $N=1$ vector  multiplet and three $N=1$ chiral multiplets. These 
supermultiplets are represented  in $N=1$ superspace  by the superfields
$V$ and $\Phi_s$ ($s=1,2,3$), which  satisfy the 
constraints $V=V^{\dag}$ and $\bar D_\dalpha \Phi_s=0$,  
$\bar D_\dalpha$ being a superspace covariant derivative\foot{We follow the same
conventions as in 
\REF\noso{J. M. F. Labastida and Carlos Lozano, ``Mathai-Quillen formulation of
Twisted $N=4$ Supersymmetric Gauge Theories in Four 
Dimensions"\journal\np&B502 (97)741; hep-th/9702106.}[\noso].}. 
The physical component 
fields of these superfields will be denoted as follows:
$$
\eqalign{V &\longrightarrow\;   A_{\alpha\dalpha},\; 
\lambda_{4\alpha},\; \bar\lambda^4{}_{\dalpha},\cr 
\Phi_s, \Phi^{\dag s} &\longrightarrow\; B_s,\;
\lambda_{s\alpha},\; B^{\dag s},\;
\bar\lambda^s{}_{\dalpha}.\cr}
\eqn\ctres
$$ 
The $N=4$ supersymmetry algebra has the automorphism group $SU(4)_I$. The field 
content of the corresponding field theory is  
conventionally arranged so that the gauge bosons are scalars under
$SU(4)_I$, while the gauginos and the scalar fields  transform 
respectively as ${\bf 4}\oplus{\bf\bar 4}$ and ${\bf 6}$. All the above fields  
take values in the adjoint representation
of some compact  Lie group $G$. 
The action   takes the following form in $N=1$ superspace:
$$
\eqalign{{\cal S} =& -{i\over 4\pi}\tau\int d^4 xd^2 \theta\, \tr (W^2) +
{i\over 4\pi}\bar\tau\int d^4 x d^2
\bar\theta\, \tr (W^{\dag 2}) \cr & 
+{1\over e^2}\sum_{s=1}^3 \int d^4 xd^2 \theta d^2
\bar\theta\, \tr(\Phi^{\dag s} \ex^V \Phi_s) \cr   
&+{i\raiz\over e^2}
\int d^4xd^2\theta \, \tr\bigl\{\Phi_1[\Phi_2 ,\Phi_3]\bigr\} +
{i\raiz\over e^2}\int d^4 xd^2\bar\theta\,\tr\bigl\{\Phi^{\dag 1} 
[\Phi^{\dag 2},\Phi^{\dag 3}]\bigr\},
 \cr }
\eqn\cuno
$$ 
where $W_\alpha =-{1\over 16}\bar D^2 \ex^{-V}D_\alpha \ex^V$ and 
$\tau={\theta\over{2\pi}}+{{4\pi^2 i}\over{e^2}}$.

The theory  is invariant under the
following four supersymmetries (in $SU(4)_I$ covariant notation):
$$
\eqalign{ 
&\delta A_{\alpha\dalpha} = -2i\bar\xi^u{}_\dalpha\lambda_{u
\alpha}+2i\bar\lambda^u{}_\dalpha\xi_{u\alpha },\cr 
&\delta\lambda_{u\alpha} =  -iF^{+}{}_{\!\alpha}{}^{\!\beta}\xi_{u\beta}+
i{\raiz}\bar\xi^{v\dot\alpha}\nabla_{\alpha\dot\alpha}
\phi_{vu}-i\xi_{w\alpha}[\phi_{uv},\phi^{vw}],\cr 
&\delta\phi_{uv}={\raiz}\bigl\{\xi_u{}^\alpha\lambda_{v\alpha}
-\xi_v{}^\alpha\lambda_{u\alpha} +
\epsilon_{uvwz}\bar\xi^w{}_{\dalpha}\bar\lambda^{z\dalpha}\bigr\},\cr}
\eqn\Vian
$$
where $F^{+}{}_{\!\alpha}{}^{\!\beta}=\sigma^{mn}{}_{\!\alpha}{}^
{\!\beta}F_{mn}$ and $(u,v,w,z,\ldots)$ label the fundamental representation ${\bf 4}$ of $SU(4)_I$. For future convenience we note that, according to our
conventions, the supersymmetry transformations with parameters 
$\xi_{v=4}^\alpha$ and $\bar\xi^{w=4}_{\dalpha}$ are the ones which are
manifest in the $N=1$ superspace formulation \cuno . In \Vian\ $\lambda_u =\{\lambda_1,\lambda_2,\lambda_3,\lambda_4\}$, while  

$$
\phi_{uv}=\!\pmatrix{\!0&-\!B^{\dag 3}&\!B^{\dag 2}&-\!B_1\cr 
                     \!B^{\dag 3}&0&\!-B^{\dag 1}&\!-B_2\cr
                     \!-B^{\dag 2}&\!B^{\dag 1}&\!0&\!-B_3\cr
                     \!B_1&        \!B_2&       \!B_3&\! 0\cr},\qquad
\cases{\phi_{uv}=-\phi_{vu}, \cr
\phi^{uv}=(\phi_{uv})^{\dag}=\phi^{*}_{vu}=
-\half\epsilon^{uvwz}\phi_{wz}\cr}
\eqn\Valle
$$ 

The global symmetry
group of $N=4$ supersymmetric theories in ${\RR}^4$ is ${\cal H}=
SU(2)_L\otimes SU(2)_R\otimes SU(4)_I$, where  ${\cal K}= 
SU(2)_L\otimes SU(2)_R$ is the rotation group $SO(4)$. The supersymmetry 
generators responsible for the transformations \Vian\ are $Q^u{}_\alpha$ and 
$\bar Q_{u\dalpha}$. They  transform  as $({\bf 2},{\bf 1},{\bf \bar 4})
\oplus({\bf 1},{\bf 2},{\bf 4})$ under ${\cal H}$.



\section{Twists of the $N=4$ theory}

Since first introduced by Witten in  [\tqft], the twisting procedure has proved
to be a very useful tool for intertwining between physical (supersymmetric)
quantum field theories and the topology of low-dimensional manifolds. In four
dimensions, the global symmetry group of the extended supersymmetric gauge
theories is of the form $SU(2)_L\otimes SU(2)_R\otimes{\cal I}$, where ${\cal
K}= SU(2)_L\otimes SU(2)_R$ is the rotation group, and ${\cal I}$ is the
chiral ${\cal R}$-symmetry group. The twist can be thought of either as an 
exotic realization of the global symmetry group of the theory, or as the 
coupling to the spin connection of a certain subgroup of the global 
${\cal R}$-current of the theory--see for 
example \REF\baryon{J. M. F. Labastida and M. Mari\~no, ``Twisted Baryon Number
in $N=2$ Supersymmetric QCD"\journal\pl&B400 (97)
323; hep-th/9702054.} [\baryon]. As
this latter mechanism changes the energy-momentum tensor and hence the
couplings (spins) of the different fields to gravity, both points of view are
easily reconciled. 

While in $N=2$ supersymmetric gauge theories 
the ${\cal R}$-symmetry group is at most $U(2)$ and
thus the twist is essentially unique, in the $N=4$ 
supersymmetric gauge theory the  
${\cal R}$-symmetry group is $SU(4)$ and there are three different
possibilities, each of these corresponding to different
non-equivalent homomorphisms of the rotation group into the ${\cal R}$-symmetry
group 
\REF\yamron{J. P. Yamron, ``Topological Actions in Twisted Supersymmetric
Theories"\journal\pl&B213 (88)325.}  
[\vafa,\noso,\yamron].  


Two of these possibilities give rise to topological field theories with two
supercharges. One of these was considered by Vafa and Witten [\vafa] in order 
to carry out an explicit test  
of $S$-duality on several four-manifolds, and is the object of the present paper. 
It has the unusual feature that the virtual dimension of
its moduli space is exactly zero. This feature was analysed from the perspective
of balanced topological field theories in 
\REF\balanced{R. Dijkgraaf and G. Moore, ``Balanced Topological Field 
Theories"\journal\cmp&185 (97)411; hep-th/9608169.}
[\balanced], while the underlying structure had already been anticipated within
 the framework of supersymmetric quantum mechanics in 
\REF\cofield{M. Blau and G. Thompson, ``$N=2$ Topological Gauge Theory, the
Euler Characteristic of Moduli Spaces, and the Casson 
Invariant"\journal\cmp&152 (93)41.}[\cofield]. 

The second possibility was first discussed in  
\REF\marcus{N. Marcus, ``The other Topological Twisting of $N=4$ Yang-Mills"\journal\np&B452 (95)331; hep-th/9506002.} [\marcus],  
where it was shown to correspond to a topological theory 
of complexified flat gauge connections.  This idea was pursued 
further in \REF\blauthomp{M. Blau and G. Thompson, ``Aspects of $N_{T}\geq 2$
Topological Gauge Theories and D-Branes"\journal\np&B492 (97)545; hep-th/9612143.} [\blauthomp], 
where a link to supersymmetric
BF-theories in four dimensions was established. From a somewhat different
viewpoint, it has been claimed in [\noso]
that the theory is amphicheiral, this meaning  
that the twist with either $SU(2)_L$ or $SU(2)_R$ leads essentially 
to the same theory. 

The remaining possibility leads to the ``half-twisted theory", 
a topological theory 
with only one BRST supercharge [\yamron]. This feature is reminiscent of the
situation in twisted $N=2$ supersymmetric gauge theories, 
and in fact [\noso], the theory is a close 
relative of the non-Abelian monopole theory 
\REF\marth{M. Mari\~no, ``The Geometry of Supersymmetric Gauge Theories 
in Four Dimensions", Ph.D. Thesis, Universidade de Santiago de Compostela, October, 1996; hep-th/9701128.}
[\marmon,\marpol,\marth], 
the non-abelian generalization of Witten's monopole theory 
 [\monop], for  the
special case in which the matter fields are in the adjoint representation of
the gauge group.


\section{The Vafa-Witten theory}

The twist of the $N=4$ supersymmetric gauge theory we are 
interested in arises as follows
[\yamron]. First break $SU(4)_I$ down to $SO(4)=SU(2)_F\otimes 
SU(2)_{F'}$, then replace $SU(2)_L$ by its diagonal sum $SU(2)'_L$ with $SU(2)_{F'}$. After the twisting, the symmetry group of the  theory becomes  ${\cal H'} =SU(2)'_L\otimes SU(2)_R\otimes SU(2)_F$. Under ${\cal H'}$, the supercharges split up as 
$$
Q^v{}_\alpha\to Q^i,\;Q^i{}_{\alpha\beta},\qquad \bar Q_{v\dalpha}\to
\bar Q^i_{\alpha\dalpha},
\eqn\elven
$$
where $i$ is an $SU(2)_F$ index. The twist has produced two scalar 
supercharges, the $SU(2)_F$ doublet $Q^i$, which are  
defined in terms of the original supercharges as follows:
$Q^{i=1}= Q^{v=1}{}_{\alpha =1} +Q^{v=2}{}_{\alpha=2}$,
$Q^{i=2}= Q^{v=3}{}_{\alpha =1} +Q^{v=4}{}_{\alpha=2}$.

The fields of the $N=4$ supersymmetric multiplet 
decompose under ${\cal H'}$ in the
following manner--in the notation of [\yamron]:
$$
\eqalign{
A_{\alpha\dalpha} &\too A_{\alpha\dalpha},\cr
\lambda_{v\alpha}&\too
\chi_{i\beta\alpha},~
\eta_{i},\cr}
\qquad\qquad
\eqalign{
\bar\lambda^v{}_{\dalpha} &\too
\psi^{i\alpha}{}_{\dalpha},\cr
\phi_{uv}&\too \varphi_{ij},~G_{\alpha\beta}.
\cr}
\eqn\poocho
$$
Notice that the fields $\chi^i_{\alpha\beta}$ and $G_{\alpha\beta}$ are
symmetric in their spinor indices and can therefore be regarded as
components of self-dual two-forms. 
\REF\BnTH{M. Blau and G. Thompson, ``Euclidean SYM Theories by Time
Reduction and Special Holonomy Manifolds", hep-th/9706225.} As argued
in [\noso] (see also [\BnTH] for a related discussion), it is convenient to 
further break  $SU(2)_F$  down to its $T_3$
subgroup, whose eigenvalues are then assumed to give  
the (non-anomalous) ghost
numbers of the different fields in the theory.  The resulting model has BRST charges $Q^{+}=Q^2$ and $Q^{-}=iQ^1$ of  opposite ghost number. The field content can now be organized
as in [\vafa],  and consists of $3$ scalar fields $\{\phi^{+2},\bar\phi^{-2},C^{0}\}$, $2$ one-forms $\{A_{\alpha\dalpha}^{0},\tilde H_{\alpha\dalpha}^{0}\}$ 
and $2$ self-dual two-forms $\{(B^{+}_{\alpha\beta})^0,
(H^{+}_{\alpha\beta})^0\}$ 
on the bosonic (commuting) side; and 
$2$ scalar fields $\{\zeta^{+1},\eta^{-1}\}$, 
$2$ one-forms $\{\psi_{\alpha\dalpha}^{1},\tilde \chi_{\alpha\dalpha}^{-1}\}$ 
and $2$ self-dual two-forms  $\{(\tilde\psi^{+}_{\alpha\beta})^{+1},
(\chi^{+}_{\alpha\beta})^{-1}\}$  
on the fermionic (anticommuting) side. The superscript stands for the ghost 
number carried by each of the fields. These fields are related to the fields 
in the underlying $N=4$ supersymmetric gauge theory as follows:

$$
\eqalign{ 
\lambda_{{\tilde  1}1}&= \tilde\psi^{+}_{11},\cr
\lambda_{{\tilde  1}2}&= \tilde\psi^{+}_{22},\cr
\lambda_{{\tilde  2}1}&= \tilde\psi^{+}_{12}-{i\over2}\zeta,\cr
\lambda_{{\tilde  2}2}&= \tilde\psi^{+}_{12}+{i\over2}\zeta,\cr
B_1&=  -B^{+}_{12}+iC,\cr 
B^{\dag}_1&=  -B^{+}_{12}-iC, 
\cr
}\quad
\eqalign{ 
\lambda_{{\tilde  3}1}&= -i\chi^{+}_{11},\cr 
\lambda_{{\tilde  3}2}&= -i\chi^{+}_{22},\cr
\lambda_{{\tilde  4}1}&= \half\eta-i\chi^{+}_{12},\cr
\lambda_{{\tilde  4}2}&= -\half\eta-i\chi^{+}_{12},\cr
B_2&= -B^{+}_{22},\cr
B^{\dag}_2&=  B^{+}_{11},  
\cr
}\quad
\eqalign{ 
\bar\lambda^{\tilde 1}_{\dalpha}&= \tilde\chi_{2\dalpha},\cr
\bar\lambda^{\tilde 2}_{\dalpha}&= -\tilde\chi_{1\dalpha},\cr
\bar\lambda^{\tilde 3}_{\dalpha}&= i\psi_{2\dot\alpha},\cr
\bar\lambda^{\tilde 4}_{\dalpha}&= -i\psi_{1\dot\alpha},\cr 
B_3&= -\bar\phi,\cr
B^{\dag}_3&= -\phi.
\cr
}
\eqn\che
$$
($\tilde 1$, $\tilde 2$, etc., denote $SU(4)_I$ indices). 


In this paper we will make use of the transformations 
generated by $Q^{+}$ only, which are readily obtained 
from \Vian\ by simply declaring  
$$
\xi_{(v=1,2)\alpha}=0,\quad
\xi_{(v=3,4)\alpha}\to \epsilon C_{(\beta=1,2)\alpha},\quad
\bar\xi^v{}_{\dalpha}=0,
\eqn\ndos
$$
and turn out to be (we give the off-shell version):

$$
\eqalign{
[Q^{+}, A_{\alpha\dalpha}] &= -2\psi_{\alpha\dalpha},\cr 
\{Q^{+},\psi_{\alpha\dot\alpha}\} &= 
-{\raiz}\deriv_{\alpha\dot\alpha}\phi, \cr
[Q^{+},\phi]&=0,\cr
[Q^{+}, B^{+}_{\alpha\beta}]&=\raiz\tilde\psi^{+}_{\alpha\beta},\cr
\{Q^{+},\tilde\psi^{+}_{\alpha\beta}\}&=2i\,[B^{+}_{\alpha\beta}, \phi],\cr
[Q^{+},C]&={1\over\raiz}\zeta,\cr
\{Q^{+},\zeta\,\} &=4i\,[C,\phi],\cr
}\qquad\quad
\eqalign{
[Q^{+},\bar\phi]&=\raiz\,\eta,\cr
\{Q^{+},\eta\,\} &=2i\,[\bar\phi,\phi],\cr
\{Q^{+},\tilde\chi_{\alpha\dalpha}\} &= 
\tilde H_{\alpha\dalpha}+\raiz s_{\alpha\dalpha},\cr
[Q^{+},\tilde H_{\alpha\dalpha}]&=
2\raiz i\,[\tilde\chi_{\alpha\dalpha},\phi]-\raiz 
[Q^{+},s_{\alpha\dalpha}]
,\cr
\{Q^{+},\chi^{+}_{\alpha\beta}\} &= H^{+}_{\alpha\beta}+
s_{\alpha\beta},\cr
[Q^{+},H^{+}_{\alpha\beta}]&=2\raiz i\,
[\chi^{+}_{\alpha\beta},\phi]-[Q^{+},s_{\alpha\beta}],\cr 
}
\eqn\papagayo
$$  
where 
$$\eqalign{ 
s_{\alpha\dalpha}&=\deriv_{\alpha\dalpha}C+
i\deriv_{\beta\dalpha}B^{+\beta}{}_\alpha,\cr
s_{\alpha\beta}&=F^{+}_{\alpha\beta}+[B^{+}_{\gamma\alpha},
B^{+}_\beta{}^{\!\gamma}]+2i\,[B^{+}_{\alpha\beta},C].\cr
}
\eqn\ecuaciones
$$
With our conventions,  the on-shell formulation is simply obtained by setting 
$H^{+}_{\alpha\beta}=0=\tilde H_{\alpha\dalpha}$ in \papagayo . 

According to Witten's fixed-point theorem  
\REF\yau{E. Witten, ``Mirror Manifolds and
Topological Field Theory",  hep-th/9112056, in {\sl Essays on Mirror Manifolds},
ed. S. T. Yau, (International  Press 1992).}
[\yau], the contributions 
to the partition function of the theory, which is the only non-trivial observable owing to the vanishing of the ghost number anomaly, 
come from the fixed points of the BRST symmetry. In view of 
\papagayo\ and \ecuaciones ,  this means 
that the Vafa-Witten theory 
localizes on the moduli space defined by the equations 
$$
\cases{ 
\deriv_{\alpha\dalpha}C+
i\deriv_{\beta\dalpha}B^{+\beta}{}_\alpha=0,\cr
F^{+}_{\alpha\beta}+[B^{+}_{\gamma\alpha},
B^{+}_\beta{}^{\!\gamma}]+2i\,[B^{+}_{\alpha\beta},C]=0,\cr
}
\eqn\masecuaciones
$$
which are precisely the equations discussed in [\vafa]. One of the main 
ingredients in the analysis in [\vafa] is the existence, on certain 
four-manifolds (basically of the K\"ahler type), of a suitable 
vanishing theorem which guarantees that all the solutions to  
 eqs. \masecuaciones\ are of the form: 

$$
F^{+}_{\alpha\beta}=0,\quad 
B^{+}_{\alpha\beta}=0, \quad C=0,
\eqn\vanishing
$$
that is, that the moduli space reduces to the moduli space of ASD connections. 
In fact, under these circumstances, the partition function of the theory 
computes, for each value of the instanton number, the Euler characteristic 
of the corresponding instanton moduli space. Observe that the vanishing 
theorem allows only positive instanton numbers to contribute to the 
partition function; the presence of negative instanton number contributions 
will signal a failure of the vanishing theorem.

\REF\wang{P. Wang, ``A Suggestion for Modification of Vafa-Witten Theory"\journal\pl&B378 (96)147; hep-th/9512021.} 
In [\vafa,\noso,\balanced,\wang] it was shown that the theory admits a nice 
geometric interpretation within the framework of the Mathai-Quillen 
formalism \REF\atiy{M. F. Atiyah and L. Jeffrey, ``Topological Lagrangians 
and Cohomology"\journal\jgp&7 (90)119.} [\atiy]
(for a review of the Mathai-Quillen formalism in the context of
topological field theories of cohomological type, see  
\REF\phyrep{D. Birmingham, M. Blau, M. Rakowski and G.
Thompson, ``Topological Field Theories", {\sl Phys. Rep.} {\bf 209} (1991),  129.}
\REF\moore{S. Cordes, G. Moore and S. Rangoolam, ``Lectures on 2D 
Yang-Mills Theory, Equivariant Cohomology and Topological Field Theory", {\sl Nucl. Phys. Proc. Suppl.} {\bf 41} (1995), 184; hep-th/9411210.}
\REF\thompson{M. Blau and G. Thompson, ``Localization and Diagonalization: a
Review of Functional Integral Techniques for Low-Dimensional Gauge Theories and
Topological Field Theories"\journal\jmp&36(93)2192; hep-th/9501075.}
[\laplata,\phyrep,\moore,\thompson]).  In this context, the equations 
\ecuaciones\ are interpreted as defining a section 
$s:{\cal M}\to {\cal V}$ in the trivial vector bundle  
${\cal V}=\mani\times{\cal F}$, where ${\cal M}={\cal A}\times\Omega^0 (X,
\ad P)\times\Omega^{2,+}(X,\ad P)$ is the field space, and the fibre is 
${\cal F}= \Omega^1(X,\ad P)\oplus\Omega^{2,+} (X,\ad P)$, whose zero locus 
--modded out by the gauge symmetry-- is precisely the desired moduli space. 
${\cal A}$ denotes the space of connections on a principal $G$-bundle $P\to
X$, while $\Omega^0(X,\ad P)$ and $\Omega^{2,+}(X,\ad P)$ denote respectively the
space of $0$-forms and self-dual $2$-forms on $X$ taking values in the Lie
algebra of $G$, while $\ad P$ denotes the adjoint bundle of $P$, $P\times_{\ad} 
{\bf g}$ (${\bf g}$ stands for the Lie algebra of $G$). The space of sections
of this bundle, $\Omega^0(X,\ad P)$, is the Lie algebra of the group ${\cal G}$ of gauge transformations (vertical automorphisms) of the bundle $P$.

In this setting, the fields of the theory play well-defined roles:
$A$, $B^{+}$ and $C$ belong to the field space; $\psi$ and $\tilde\psi^{+}$ 
are ghosts living in the (co)tangent space $T^*{\cal M}$; 
$\tilde\chi$ and $\chi^{+}$ are fibre antighosts associated to eqs.  
\ecuaciones , while $\tilde H$ 
and $H^{+}$ are their corresponding auxiliary fields; finally, $\phi$ --or 
rather its vacuum expectation value $\langle\phi\rangle$-- gives the curvature 
of the principal 
${\cal G}$-bundle ${\cal M}\to {\cal M}/{\cal G}$, while $\bar\phi$ and $\eta$ 
enforce the horizontal projection ${\cal M}\to {\cal M}/{\cal G}$. The BRST 
symmetry 
\papagayo\ is the Cartan model representative of the ${\cal G}$-equivariant 
differential on ${\cal V}$, 
while the ghost number is just a form degree. The exponential of the action of 
the theory gives, when integrated over the antighosts and their auxiliary
fields, the Mathai-Quillen representative for the Thom form of the principal 
bundle ${\cal M}\times{\cal F}\to {\cal E}={\cal M}\times_{\cal G} {\cal F}$.  

The action itself (but for the theta-term) can be written as a 
$Q^{+}$ commutator. 
The appropriate gauge fermion is [\noso]:
$$
\eqalign{
\Psi&=
{1\over e^2}\int_X  d^4 x\,\sqrt{g}\, \tr
\left\{\,-{1\over4}
\tilde \chi^{\dalpha\alpha}\bigl (\,\tilde H_{\alpha\dalpha}-\raiz 
s_{\alpha\dalpha}\bigr )-{1\over 4} \chi^{\alpha\beta}\bigl (\, 
H_{\alpha\beta}-s_{\alpha\beta}\bigr)\,\right\}\cr 
&+{1\over e^2}\int_X  d^4 x\,\sqrt{g}\, \tr
\left\{\,
{1\over{2\raiz}}\bar\phi\,\bigl
(\,\deriv_{\alpha\dalpha}\psi^{\dot\alpha\alpha}+i\raiz
\,[\tilde\psi_{\alpha\beta},B^{\alpha\beta}]
-i\raiz\,[\zeta,C]\,\bigr )\,\right\}\cr
&-{1\over e^2}\int_X  d^4 x\,\sqrt{g}\, \tr
\left\{\,{i\over4}\eta[\phi,\bar\phi]\,\right\}.\cr
}
\eqn\Mazinger
$$

We have not said a word about the role played by $Q^{-}$. In fact, 
the theory admits two   Mathai-Quillen descriptions, related to each other by 
the Weyl group of $SU(2)_F$, 
in such a way that the roles of $Q^{+}$ and $Q^{-}$ are interchanged, 
as are the roles of $\psi$ and $\tilde\chi$, $\chi^{+}$ and $\tilde\psi^{+}$, 
$\zeta$ and $\eta$, and $\phi$ and $\bar\phi$. The corresponding moduli space 
is defined by eqs. \masecuaciones\ with the 
substitution $C\to -C$, and the 
theory  localizes --as was proved in [\vafa]-- actually on the intersection of 
both 
moduli spaces, which is defined by the equations 
$$
\cases{ 
\deriv_{\alpha\dalpha}C=0,\qquad 
\deriv_{\beta\dalpha}B^{+\beta}{}_\alpha=0,\cr
F^{+}_{\alpha\beta}+[B^{+}_{\gamma\alpha},
B^{+}_\beta{}^{\!\gamma}]=0,\quad [B^{+}_{\alpha\beta},C]=0.\cr
}
\eqn\ecuacions
$$


\section{The twist on K\"ahler manifolds}

On a four-dimensional K\"ahler manifold the holonomy group is contained in
$SU(2)_R\otimes U(1)_L$, where $U(1)_L$ is a certain subgroup of $SU(2)_L$.
Under this reduction of the holonomy, left-handed spinors $\psi_\alpha$
decompose into pieces $\psi_1$ and $\psi_2$ of opposite  $U(1)_L$ charges, in
such a way that if the manifold is also spin, the spinor bundle $S^+$ has a
decomposition $S^+\simeq K^\half \oplus  K^{-\half}$, where  $K^\half$ is some
square root of the canonical bundle of $X$, $K=\bigwedge^2_{\bf {C}}T^{*}X$. 
We can define a complex structure on $X$ by taking the $1$-forms
$(\sigma_\mu)_{1\dalpha}dx^\mu$ to be of type $(1,0)$, and the $1$-forms    
$(\sigma_\mu)_{2\dalpha}dx^\mu$ of type $(0,1)$. With this choice, the
self-dual $2$-form $(\sigma_{\mu\nu})_{\alpha\beta}dx^\mu\wedge dx^\nu$ can be
regarded as a $(2,0)$-form for $\alpha=\beta=1$, as a $(0,2)$-form
for $\alpha=\beta=2$, and as a $(1,1)$-form for $\alpha=1, \beta=2$. This
decomposition corresponds to the splitting 
$\Omega^{2,+}(X)=\Omega^{2,0}(X)\oplus\Omega^{0,2}(X)\oplus
\varpi\Omega^{0}(X)$, valid on any K\"ahler surface ($\varpi$ stands for the
K\"ahler form).  

With respect to the complex structure of the manifold, the fields of the theory
 naturally split into objects that can be thought of as components of forms of
type $(p,q)$. For example, the connection $1$-form $A_{\alpha\dalpha}
(\sigma_\mu)^{\dalpha\alpha}dx^\mu$  splits up into a $(1,0)$-form 
$A_{2\dalpha}(\sigma_\mu)^{\dalpha}_1 dx^\mu$ and a $(0,1)$-form 
$A_{1\dalpha}(\sigma_\mu)^{\dalpha}_2 dx^\mu$. Likewise, the self-dual $2$-form 
$B^{+}_{\alpha\beta}(\sigma_{\mu\nu})^{\alpha\beta}dx^\mu\wedge dx^\nu$ gives
rise to a $(2,0)$-form  
$B^{+}_{22}(\sigma_{\mu\nu})_{11}dx^\mu\wedge dx^\nu$
a $(0,2)$-form
$B^{+}_{11}(\sigma_{\mu\nu})_{22}dx^\mu\wedge dx^\nu$
and  a $(1,1)$-form for 
$B^{+}_{12}(\sigma_{\mu\nu})_{12}dx^\mu\wedge dx^\nu=B^{+}_{12}\varpi$. 
Notice that in our conventions the field $B^{+}_{11}$ would correspond 
to the $(0,2)$-form $\bar\beta$, $B^{+}_{22}$  
to the $(2,0)$-form $\beta$ and  $B^{+}_{12}$ to the $0$-form $b$ in 
[\vafa]. Note
that the field $B^{+}_{12}$ can be thought of as a scalar field on $X$. In
fact, we shall see in a moment that it naturally combines with the scalar
field $C$ into two complex scalars $B^{+}_{12}\pm iC$. Something similar
happens with the other self-dual $2$-forms $\chi^{+}$ and $\tilde\psi^{+}$. 

Let us recall that in our conventions the BRST operators $Q^{\pm}$ are 
obtained 
from the $N=4$ supercharges $Q^v{}_\alpha$, with the recipe
$$
Q^{+}=Q^{\tilde 3}{}_1+Q^{\tilde 4}{}_2,\qquad 
Q^{-}=i(Q^{\tilde 1}{}_1+Q^{\tilde 2}{}_2).
\eqn\rabindranah
$$
In the K\"ahler case, each of the individual components $Q^{\tilde 1}{}_1$, 
$Q^{\tilde 2}{}_2$, $Q^{\tilde 3}{}_1$ and $Q^{\tilde 4}{}_2$ is  well-defined
under the holonomy $SU(2)_R\otimes U(1)_L$. It is therefore possible to define
four charges, of which only   $Q^{\tilde 4}{}_2$ is related to the underlying
construction  in $N=1$ superspace. Hence, it is the only topological symmetry
that should be expected to survive after the mass terms are plugged in. 
\endpage
In what
follows, we will be interested only in $Q^{\tilde 3}{}_1$ and $Q^{\tilde
4}{}_2$. The corresponding transformation laws (with parameters $\rho_2$ and  
$\rho_1$ respectively) can be extracted from the $N=4$ supersymmetry 
transformations \Vian\ by setting: 
$$
\bar\xi^{v\dalpha}=0,\quad \xi_{\tilde 1\alpha}= 
\xi_{\tilde 2\alpha}=0,\quad
\xi_{\tilde 3\alpha}=\rho_2 C_{2\alpha}, \quad 
\xi_{\tilde 4\alpha}=\rho_1 C_{1\alpha},
\eqn\tagore      
$$
The corresponding BRST charges will be denoted by $Q_1=Q^{\tilde 4}{}_2$
and $Q_2=Q^{\tilde 3}{}_1$. The on-shell transformations turn out to be:
$$
\eqalign{ 
[Q_1,A_{1\dot\alpha}] &= -2\psi_{1\dot\alpha},\cr 
[Q_1,A_{2\dot\alpha}] &=0,\cr
[Q_1,F^{+}_{11}]&=-2iD_{1\dalpha}\psi_1{}^\dalpha,\cr
[Q_1,F^{+}_{22}]&=0,\cr
\{Q_1,\psi_{1\dot\alpha}\}&=0,\cr 
\{Q_1,\psi_{2\dot\alpha}\}&=-\raiz D_{2\dot\alpha}\phi,\cr
[Q_1,\phi]&=0, \cr
[Q_1, B^{+}_{11}] &= 0,\cr 
[Q_1, B^{+}_{12}+iC] &=0 ,\cr
\{Q_1, \tilde\psi^{+}_{11}\}&=2i[ B^{+}_{11},\phi],\cr
\left\{Q_1,\tilde\psi^{+}_{12}+{i\over2}\zeta\right\}&
=-2i[\phi,B^{+}_{12}+iC],\cr
\{Q_1, \tilde\chi_{1\dalpha}\}&=-\raiz iD_{2\dalpha} B^{+}_{11},\cr
}
\qquad\quad
\eqalign{
[Q_1,F^{+}_{12}]&=-iD_{2\dalpha}\psi_1{}^\dalpha,\cr
[Q_1, \bar\phi] &= \raiz\left(\half\eta-i\chi^{+}_{12}\right),\cr 
\left\{Q_1, \half\eta +i\chi^{+}_{12}\right\} &=  -i[\phi,\bar\phi]+iF^{+}_{12}\cr
+{i}[B^{+}_{12}-iC,B^{+}_{12}+iC]+&{i}[B^{+}_{11}, B^{+}_{22}],\cr
\left\{Q_1, \half\eta-i\chi^{+}_{12}\right\}&= 0,\cr
\{Q_1,\chi^{+}_{11} \}&= -2[B^{+}_{12}+iC,B^{+}_{11}],\cr
\{Q_1,\chi^{+}_{22} \}&= F^{+}_{22},\cr
[Q_1, B^{+}_{22}] &= \raiz\tilde\psi^{+}_{22},\cr 
[Q_1, B^{+}_{12}-iC] &=\raiz\left(\tilde\psi^{+}_{12}-{i\over2}
\zeta\right) ,\cr
\{Q_1, \tilde\psi^{+}_{22}\}&=0,\cr
\left\{Q_1, \tilde\psi^{+}_{12}-{i\over2}\zeta\right\}&=0,\cr
\{Q_1, \tilde\chi_{2\dalpha}\}&=-\raiz iD_{2\dalpha}(B^{+}_{12}+iC),\cr
}
\eqn\sandia
$$
for $Q_1$. The $Q_2$ transformations are easily computed from \papagayo\ and 
\sandia\  after using $Q^{+}=Q_1+Q_2$. They read:

$$
\eqalign{ 
[Q_2,A_{1\dot\alpha}] &=0,\cr
[Q_2,A_{2\dot\alpha}] &= -2\psi_{2\dot\alpha},\cr
[Q_2,F^{+}_{11}]&=0,\cr
[Q_2,F^{+}_{22}]&=-2iD_{2\dalpha}\psi_2{}^\dalpha,\cr\cr
\{Q_2,\psi_{1\dot\alpha}\}&=-\raiz D_{1\dot\alpha}\phi,\cr
\{Q_2,\psi_{2\dot\alpha}\}&=0,\cr 
[Q_2,\phi]&=0, \cr
[Q_2, B^{+}_{11}] &= \raiz\tilde\psi^{+}_{11},\cr  
[Q_2, B^{+}_{12}-iC] &=0 ,\cr
\{Q_2, \tilde\psi^{+}_{11}\}&=0,\cr
\left\{Q_2, \tilde\psi^{+}_{12}-{i\over2}\zeta\right\}&=-2i[\phi,B^{+}_{12}-iC],\cr
\{Q_2, \tilde\chi_{1\dalpha}\}&=\raiz iD_{1\dalpha}(B^{+}_{12}-iC),\cr
}
\qquad\quad
\eqalign{
[Q_2,F^{+}_{12}]&=-iD_{1\dalpha}\psi_2{}^\dalpha,\cr
[Q_2, \bar\phi] &= \raiz\left(\half\eta+i\chi^{+}_{12}\right),\cr 
\left\{Q_2, \half\eta-i\chi^{+}_{12}\right\} &=  -i[\phi,\bar\phi]-iF^{+}_{12}\cr
-{i}[B^{+}_{12}-iC,B^{+}_{12}+iC]&-{i}[B^{+}_{11}, B^{+}_{22}],\cr
\left\{Q_2, \half\eta+i\chi^{+}_{12}\right\}&= 0,\cr
\{Q_2,\chi^{+}_{11} \}&= F^{+}_{11},\cr
\{Q_2,\chi^{+}_{22} \}&= 2[B^{+}_{12}-iC,B^{+}_{22}],\cr
[Q_2, B^{+}_{22}] &= 0,\cr 
[Q_2, B^{+}_{12}+iC] &=\raiz\left(\tilde\psi^{+}_{12}+{i\over2}\zeta\right) ,\cr
\{Q_2, \tilde\psi^{+}_{22}\}&=2i[ B^{+}_{22},\phi],\cr
\{Q_2,\tilde\psi^{+}_{12}+{i\over2}\zeta\}&=0,\cr
\{Q_2, \tilde\chi_{2\dalpha}\}&=\raiz iD_{1\dalpha} B^{+}_{22},\cr
}
\eqn\pavia
$$
It is straightforward to verify that $(Q_1)^2=(Q_2)^2=0$ on-shell, while 
$\{Q_1,Q_2\}$ gives a gauge transformation generated by $\phi$.  

\vfill\endpage


\chapter{Mass perturbations}

We now turn to the discussion of the possible ways of (softly) breaking $N=4$
supersymmetry by suitably adding mass terms for the chiral multiplets. Let us
 consider first the situation that arises on a flat ${\RR}^4$. By adding a
bare mass term for just one of the chiral multiplets, say $\Phi_1$, 
$$
\Delta L^{(1)} = m\int d^4 xd^2\theta\tr{(\Phi_1)^2}+{\hbox{\rm h.c.}}, 
\eqn\pertu
$$
$N=4$ supersymmetry is broken down to $N=1$. The corresponding low-energy
effective theory, at scales below $m$, is $N=1$ supersymmetric  QCD, with $SU(2)$ as gauge group, coupled to two 
massless chiral superfields in the adjoint representation with a (tree-level) 
quartic superpotential induced by integrating out the massive superfield. As 
shown in \REF\phases{K. Intriligator and
N. Seiberg, ``Phases of $N=1$ Supersymmetric Gauge Theories in Four 
Dimensions"\journal\np&B431 (94)551; hep-th/9408155.} [\phases], 
this theory has a moduli space 
of vacua where both a Coulomb and a Higgs phase coexist. 
On 
the other hand, equal bare mass terms for two of the chiral multiplets, 
$$
\Delta L^{(2)} = m\int d^4 xd^2\theta\tr{(\Phi_1\Phi_2)}+{\hbox{\rm h.c.}}, 
\eqn\pertur
$$
preserve $N=2$ supersymmetry, whereas if the mass terms are different: 
$$  
{\Delta}' L^{(2)} = m_1\int d^4 xd^2\theta\tr{(\Phi_1)^2}+
m_2\int d^4 xd^2\theta\tr{(\Phi_2)^2}+{\hbox{\rm h.c.}}, 
\eqn\perturb
$$
$N=4$ supersymmetry is again broken down to $N=1$. However, both theories flow in the
infrared to a pure $N=2$ supersymmetric gauge theory, which has a moduli space of vacua  
in the Coulomb phase. 
Finally, mass terms for 
the three 
chiral multiplets, no matter whether the mass parameters are equal or not,
preserve only $N=1$ supersymmetry. Of the three inequivalent ways of breaking
$N=4$ supersymmetry down to $N=1$, we must choose the one in terms of which the analysis of
the vacuum structure of the resultant $N=1$ theory is simplest. The 
appropriate choice is [\vafa] 
$$
\Delta L^{(3)} = m\int d^4 xd^2\theta\tr{\bigl((\Phi_1)^2+(\Phi_2)^2+(\Phi_3)^2
\bigr)}
+{\hbox{\rm h.c.}}, 
\eqn\perturba
$$
in terms of which the classical vacua of the resulting $N=1$ theory can be
classified by the complex conjugacy classes of homomorphisms of the $SU(2)$ Lie
algebra to that of $G$. In the case that $G=SU(2)$, there are three discrete
vacua, corresponding to the three singularities of the mass-deformed $N=4$  
 supersymmetric gauge theory with gauge group $SU(2)$ [\swotro].     

On general curved manifolds the na{\"\i}ve construction sketched above simply
doesn't work. As explained in [\wijmp,\vafa], superpotentials of a twisted
theory on K\"ahler manifolds must transform as $(2,0)$-forms. According to our
conventions, two of the chiral superfields, $\Phi_1$ and $\Phi_3$ (whose scalar
components are $B^{+}_{12}\pm iC$ and $\phi$, $\bar\phi$ resp.) are scalars in
the twisted model, while the third one, $\Phi_2$ (whose scalar
components are $B^{+}_{11}$ and  $B^{+}_{22}$), is a $(2,0)$-form. A suitable 
mass term for $\Phi_2$ and one of the other scalar superfields, say $\Phi_1$,
can be readily written down and reads:

$$
\Delta L(m) = m\int_X d^2\theta\tr{(\Phi_1\Phi_2)}+{\hbox{\rm h.c.}}
\eqn\perturm
$$
In \perturm\ $m$ is just a (constant) mass parameter. A mass term for the
remaining superfield $\Phi_3$ requires the introduction of the $(2,0)$-form
\foot{Of course, this sets on the manifold $X$ the constraint 
$h^{(2,0)}(X)\not=0$, which for K\"ahler manifolds is equivalent to 
demanding $b^{+}_2>1$. This excludes, for example, the case of $\CP^2$.} 
$\omega$ [\wijmp]:
$$
\Delta L(\omega) = \int_X
\omega\wedge d^2{\bar z}d^2\theta\tr{(\Phi_3)^2}+{\hbox{\rm h.c.}}
\eqn\perturbation
$$

Therefore we now turn to studying the effect of the following mass terms for 
the chiral multiplets $\Phi_1$, $\Phi_2$ and  $\Phi_3$:
$$\eqalign{
\Delta L (m,\omega)&= m\int_X d^2\theta\tr{(\Phi_1\Phi_2)}+
m\int_X d^2\bar\theta\tr{(\Phi^{\dag}_1\Phi^{\dag}_2)}\cr
+&\int_X d^2\theta\omega\tr{(\Phi_3)^2}+
\int_X d^2\bar\theta\bar\omega\tr{(\Phi^{\dag}_3)^2},
\cr}
\eqn\pera
$$
where, for simplicity, $\omega=\omega_{11}=(\sigma_{\mu\nu})_{11}\omega_{\tau\lambda}\epsilon^{\mu\nu
\tau\lambda}$ stands for the only non-vanishing component of the $(2,0)$-form 
$\omega$, while $\bar\omega=\bar\omega_{22}=\omega_{11}^{*}$ stands for the
only non-vanishing component of the $(0,2)$-form $\bar\omega$ conjugate to 
$\omega$. 

After expanding the fields and integrating out the auxiliary fields one gets 
the contributions
$$
\eqalign{
-&2\raiz i \omega B_3[B^{\dag}_1, B^{\dag}_2]-2\raiz i \bar\omega 
B^{\dag}_3[B_1, B_2]-4\vert\omega\vert^2 B_3 B^{\dag}_3\cr
-&\omega\lambda_3{}^{\alpha} \lambda_{3\alpha}-\bar\omega
\bar\lambda^3{}_{\dalpha} \bar\lambda^{3\dalpha}\cr
-&2\raiz i m B_2[B^{\dag}_2, B^{\dag}_3]-2\raiz i m 
B^{\dag}_2[B_2, B_3]-m^2 B_2 B^{\dag}_2\cr
-&m\lambda_1{}^{\alpha} \lambda_{2\alpha}-m
\bar\lambda^1{}_{\dalpha} \bar\lambda^{2\dalpha}\cr
-&2\raiz i m B_1[B^{\dag}_3, B^{\dag}_1]-2\raiz i m 
B^{\dag}_1[B_3, B_1]-m^2 B_1 B^{\dag}_1.\cr}
\eqn\manzana
$$
The $N=1$ transformations for the fermions get modified as follows:
$$
\eqalign{
\delta \lambda_{1\alpha}&=\ldots -\raiz\xi_{4\alpha}mB^{\dag}_2,\cr
\delta \lambda_{2\alpha}&=\ldots -\raiz\xi_{4\alpha}mB^{\dag}_1,\cr
\delta \lambda_{3\alpha}&=\ldots -2\raiz\xi_{4\alpha}\bar\omega 
B^{\dag}_3\cr}
\eqn\peach
$$
(and their corresponding complex conjugates). 
In terms of the twisted fields the mass contributions are --see \che :

$$
\eqalign{
\tr &\Bigl\{-2\raiz i\omega\bar\phi[B^{+}_{12}+iC,B^{+}_{11}]+ 2\raiz 
i\bar\omega
\phi[B^{+}_{12}-iC,B^{+}_{22}]-4\vert\omega\vert^2\phi\bar\phi\cr
-&2i\omega\chi^{+}_{11}\left(\half\eta-i\chi^{+}_{12}\right)+
\bar\omega\psi_{2\dalpha}
\psi_2{}^{\dalpha}-2\raiz imB^{+}_{22}[B^{+}_{11},\phi]\cr
-&2\raiz imB^{+}_{11}[B^{+}_{22},\bar\phi]+m^2 B^{+}_{11}B^{+}_{22}+ m
\left(\tilde\psi^{+}_{12}+{i\over2}\zeta\right)\left(\tilde\psi^{+}_{12}-
{i\over2}\zeta\right)
\cr+&m\tilde\psi^{+}_{11}\tilde\psi^{+}_{22}
+m\tilde\chi_{2\dalpha}\tilde\chi_1{}^\dalpha +
\raiz im \phi[B^{+}_{12}+iC,B^{+}_{12}-iC]\cr-&\raiz im \bar\phi
[B^{+}_{12}+iC,B^{+}_{12}-iC]- m^2 \vert B^{+}_{12}+iC\vert^2\Bigr\}.\cr
}
\eqn\uva
$$
Notice that the mass terms \uva\ explicitly break the ghost number symmetry. In
fact, as there are terms with ghost number $+2$, others with ghost number $-2$,
and finally some with ghost number $0$, the perturbation actually preserves a 
${\ZZ}_2$ subgroup of the ghost number. 
The $Q_1$ transformations \sandia , which are the only ones to survive the perturbation a priori, also get modified in a way that is dictated
by the underlying $N=1$ structure, so that in view of  \peach\ they become:
 
$$
\eqalign{ 
\{Q^{(m,\omega)}_1, \tilde\psi^{+}_{11}\}&=2i[ B^{+}_{11},\phi]-
\raiz m B^{+}_{11},\cr
\left\{Q^{(m,\omega)}_1,\tilde\psi^{+}_{12}+{i\over2}\zeta\right\}
&=-2i[\phi,B^{+}_{12}+iC]
+\raiz m(B^{+}_{12}+iC),\cr
\{Q^{(m,\omega)}_1,\chi^{+}_{11} \}&= -2[B^{+}_{12}+iC,B^{+}_{11}]
+2\raiz i\bar\omega\phi.\cr
}                                                                        
\eqn\sandy
$$
(The rest of the transformations remain the same.) 
Notice that the fixed-point equations which stem from \sandy\ are precisely   
the $F$-flatness conditions as derived from the superpotential 
$$
i\raiz \tr\left(\Phi_1[\Phi_2,\Phi_3]\right)+m\tr\left(\Phi_1\Phi_2\right)
+\omega\tr(\Phi_3)^2.
\eqn\sakharov
$$
We can analyse these equations following [\vafa]. They admit a trivial 
solution $B^{+}_{11}=B^{+}_{12}=C=\phi=0$, which leaves at low energies the
two vacua of the pure $N=1$ supersymmetric gauge theory with gauge group $SU(2)$. Unless the manifold $X$ 
is hyper-K\"ahler, this picture must be corrected near the zeroes of the
mass parameter $\omega$ (which form a collection of complex one-dimensional
submanifolds $\{C_i\}$) along the lines proposed in [\wijmp]. In addition to
this trivial vacuum, eqs. \sandy\ admit a non-trivial fixed-point in
which $\phi$, and therefore $B^{+}_{11}$, $B^{+}_{12}$ and $C$, are not zero. 
On flat space-time this solution corresponds to a Higgs vacuum in which the 
gauge group is completely broken. From the viewpoint of the mass-perturbed 
$N=2$ supersymmetric gauge theory, it corresponds, at least 
for large $m$, to a singular point where 
an elementary quark hypermultiplet becomes massless [\swotro]. This analysis is
still valid on hyper-K\"ahler manifolds. However, on arbitrary K\"ahler
manifolds, this vacuum bifurcates into a ``Higgs" vacuum where the gauge 
group is completely broken, and an Abelianised vacuum with gauge bundle 
$E=K^{1/2}\oplus K^{-1/2}$ and instanton number 
$n=-{{2\chi+3\sigma}\over4}$. This Abelianisation can be understood as follows 
[\vafa]. On K\"ahler manifolds eqs. \ecuaciones\ can be decomposed in
the following way (this can seen by looking at the 
$Q_{1,2}$-transformations \sandia , \pavia ):
$$
\eqalign
{
F^{+}_{11}&=0=F^{+}_{22},\qquad [B^{+}_{12}+ iC, B^{+}_{11}]=0=[B^{+}_{12}-iC,
B^{+}_{22}],\cr  
F^{+}_{12}&+[B^{+}_{12}-iC,B^{+}_{12}+iC]+[B^{+}_{11},B^{+}_{22}]=0.\cr
}
\eqn\nastasha
$$
These equations have a $U(1)$ symmetry (which will be further exploited below) 
$B^{+}_{11}\to \ex^{i\alpha}B^{+}_{11}$, $B^{+}_{22}\to
\ex^{-i\alpha}B^{+}_{22}$, $B^{+}_{12}\pm iC \to  \ex^{\mp i\alpha}(B^{+}_{12}
\pm iC)$. When the vanishing theorem fails, the contributions from the branch 
$B^{+}\not = 0\not = C$ come from the fixed points of the combined 
gauge-$U(1)$ action. If there is a non-trivial fixed point,  
the gauge connection has to be reducible there, and the gauge bundle is therefore Abelianised. The instanton number of such an Abelianised bundle is typically negative, which means that on a general K\"ahler manifold, the partition function of the theory will be computing not the Euler
characteristic of the instanton moduli space (recall that the contribution of bundles with negative instanton number means that the vanishing theorem is failing), but the Euler characteristic of
the $U(1)$-equivariant bundle defined by eqs. \nastasha .    

With the mass terms added, the action $ S+\Delta L(m,\omega)$ is only 
invariant under $Q^{(m,\omega)}_1$.  To get rid of the mass terms
proportional to $m$, we shall proceed as follows. We will modify the
$Q_2$ transformations by appropriately introducing mass terms (proportional to
$m$), in such a way that $Q^{+}(m)=Q^{(m)}_1+Q^{(m)}_2$ (with mass $m$, and 
$\omega=0$ at this stage)  be a symmetry of the original action plus mass
perturbations. We will show this by proving that $L+\Delta L(m,\omega=0)$ is
actually $Q^{+}(m)$-exact. To this end we make the replacements:
$$
\eqalign{
\{Q_2, \tilde\psi^{+}_{22}\}=2i[ B^{+}_{22},\phi]&\too 
\{Q^{(m)}_2, \tilde\psi^{+}_{22}\}=2i[ B^{+}_{22},\phi]+\raiz mB^{+}_{22}
,\cr
\left\{Q_2, \tilde\psi^{+}_{12}-{i\over2}\zeta\right\}=-2i[\phi,B^{+}_{12}-iC]&\too
\left\{Q^{(m)}_2,\tilde\psi^{+}_{12}-{i\over2}\zeta\right\}\cr 
&\quad =-2i[\phi,B^{+}_{12}-iC] -2\raiz m(B^{+}_{12}-iC)
\cr}
\eqn\mesa
$$
(the rest of the transformations remain the same). 
Notice that still $(Q^{(m)}_2)^2=0$. Next we spell out the
$Q^{+}(m)=Q^{(m)}_1+Q^{(m)}_2$-transformations: 

$$ 
\eqalign{
[Q^{+}(m), B^{+}_{11}] &= \raiz\tilde\psi^{+}_{11},\cr 
[Q^{+}(m), B^{+}_{22}] &= \raiz\tilde\psi^{+}_{22},\cr 
[Q^{+}(m), B^{+}_{12}\pm iC] &=\raiz\left(\tilde\psi^{+}_{12}\pm {i\over2}\zeta\right),\cr
\left\{Q^{+}(m), \tilde\psi^{+}_{12}\pm {i\over2}\zeta\right\}&=
2i[B^{+}_{12}\pm iC,\phi]\cr
&\pm\raiz m (B^{+}_{12}\pm iC),\cr
}
\qquad
\eqalign{
\{Q^{+}(m), \tilde\psi^{+}_{11}\}&=2i[ B^{+}_{11},\phi]-\raiz m B^{+}_{11},\cr
\{Q^{+}(m), \tilde\psi^{+}_{22}\}&=2i[ B^{+}_{22},\phi]+\raiz m B^{+}_{22},\cr
}
\eqn\silla
$$
On any of these fields (which we denote generically by $X$) 
the charge $Q^{+}(m)$ satisfies the algebra:
$$
(Q^{+}(m))^2 X=2\raiz i[X,\phi]+2mqX, 
\eqn\equivariant
$$
where  $q =-1$  for $B^{+}_{11}$, $\tilde\psi^{+}_{11}$,  
$B^{+}_{12}-iC$ and $\tilde\psi^{+}_{12}-{i\over2}\zeta$, and  $q =+1$
for $B^{+}_{22}$, $\tilde\psi^{+}_{22}$,  
$B^{+}_{12}+iC$ and $\tilde\psi^{+}_{12}+{i\over2}\zeta$. 
Notice that these charge assingments  are compatible with the $U(1)$ symmetry
that we discussed above, and in fact one can see the ``central charge" 
$\delta_q X=2mqX$ arising in the algebra \equivariant\ as an infinitesimal 
$U(1)$ transformation with parameter $m$. 

We also extend the $Q^{+}(m)$ transformation off-shell by declaring  
its action on $\tilde H_{\alpha\dalpha}$ to be:
$$
\eqalign{ 
[Q^{+}(m),\tilde H_{1\dalpha}]&=\ldots -2m\tilde\chi_{1\dalpha},\cr
[Q^{+}(m),\tilde H_{2\dalpha}]&=\ldots +2m\tilde\chi_{2\dalpha}.\cr 
}
\eqn\papaya
$$
In this way, $Q^{+}(m)$ closes on $\tilde H_{1\dalpha},\tilde\chi_{1\dalpha}$ 
with $q=-1$, and on $\tilde H_{2\dalpha},\tilde\chi_{2\dalpha}$ with $q=+1$.
 
Let us now prove that the above modifications suffice to render the $m$ mass
terms $Q^{+}(m)$-exact:   
$$
{1\over{2\raiz}}m(\tilde\psi^{+}_{22}B^{+}_{11}-\tilde\psi^{+}_{11}B^{+}_{22})
\buildrel Q^{+}(m)\over \longrightarrow m^2 B^{+}_{11}B^{+}_{22}+
m\tilde\psi^{+}_{11}\tilde\psi^{+}_{22}-\raiz imB^{+}_{22}[B^{+}_{11},\phi],
\eqn\Rimbaud
$$
and 
$$
\eqalign{
-{1\over{2\raiz}}&m\left\{(B^{+}_{12}-iC)\left(\tilde\psi^{+}_{12}+{i\over2}\zeta\right)-
(B^{+}_{12}+iC)\left(\tilde\psi^{+}_{12}-{i\over2}\zeta\right)\right\}
\buildrel Q^{+}(m)\over \longrightarrow \cr
&-m^2 \vert B^{+}_{12}+iC\vert^2+\raiz im\phi[B^{+}_{12}+iC,B^{+}_{12}-iC]
+m\left(\tilde\psi^{+}_{12}+{i\over2}\zeta\right)\left(\tilde\psi^{+}_{12}-
{i\over2}\zeta\right). 
\cr}
\eqn\Apollinaire
$$
Notice, moreover, that these terms are likewise   
$Q^{(m,\omega)}_{1}$-exact:
$$
-{1\over{\raiz}}m\tilde\psi^{+}_{11}B^{+}_{22}
\,\buildrel Q^{(m,\omega)}_1\over \longrightarrow m^2 B^{+}_{11}B^{+}_{22}+
m\tilde\psi^{+}_{11}\tilde\psi^{+}_{22}-\raiz imB^{+}_{22}[B^{+}_{11},\phi],
\eqn\isa
$$
and
$$
\eqalign
{
-\raiz m\left\{(B^{+}_{12}-iC)\tilde\psi^{+}_{12}\right\}\,
\buildrel Q^{(m,\omega)}_1\over \longrightarrow 
-m^2 \vert B^{+}_{12}+iC\vert^2&+\raiz im\phi[B^{+}_{12}+iC,B^{+}_{12}-iC]
\cr&+m\left(\tilde\psi^{+}_{12}+{i\over2}\zeta\right)\left(\tilde
\psi^{+}_{12}-{i\over2}\zeta\right). 
\cr}
\eqn\ennes
$$
But we have not yet reproduced the terms (see \uva):
$-\raiz im\bar\phi[B^{+}_{12}+iC,B^{+}_{12}-iC]$, 
$-2\raiz imB^{+}_{11}[B^{+}_{22},\bar\phi]$
and $m\tilde\chi_{2\dalpha}\tilde\chi_1{}^\dalpha$. These come from pieces 
already present in the gauge fermion. Explicitly,  
$$
 \tr
\left\{-{1\over4}
\tilde \chi^{\dalpha\alpha}\tilde H_{\alpha\dalpha}\right\} 
\buildrel Q^{+}(m)\over \longrightarrow 
m\tilde\chi_{2\dalpha}\tilde\chi_1{}^\dalpha ,
\eqn\sumathi
$$
and 

$$
\eqalign{
\tr &\left\{
{i\over{2}}\bar\phi[\tilde\psi_{\alpha\beta},B^{\alpha\beta}]
+{i\over{2}}[\zeta,C]\right\}\buildrel Q^{+}(m) \over \longrightarrow \cr
&-\raiz im\bar\phi[B^{+}_{12}+iC,B^{+}_{12}-iC] 
-2\raiz imB^{+}_{11}[B^{+}_{22},\bar\phi].\cr
}
\eqn\fujita
$$

The analysis of the terms containing the $(2,0)$-form $\omega$ can be carried
out essentialy as in the Donaldson-Witten theory. The perturbation breaks up into a
$ Q^{(m,\omega)}_1$-exact piece:

$$
\eqalign{
\{ Q^{(m,\omega)}_1,\tr(\raiz i\omega\bar\phi\chi^{+}_{11})\}&\cr
=\tr\Biggl\{-2\raiz i&\omega\bar\phi[B^{+}_{12}+iC,B^{+}_{11}] 
-4\vert\omega\vert^2\phi\bar\phi
-2i\omega\chi^{+}_{11}\left(\half\eta-i\chi^{+}_{12}\right)\Biggr\},\cr}
\eqn\pinha
$$
and an operator of ghost number $+2$:
$$
J(\bar\omega)=\int_X \tr\left(2\raiz i\bar\omega
\phi[B^{+}_{12}-iC,B^{+}_{22}]+\bar\omega\psi_{2\dalpha}
\psi_2{}^{\dalpha}\right).
\eqn\cereza
$$ 
Equation \cereza\ is not very useful as it stands. To rewrite it in a more convenient
form we note that from \papagayo\ it follows that:
$$
2\raiz i\bar\omega\tr\left\{\phi[B^{+}_{12}-iC,B^{+}_{22}]\right\}=
\raiz i\tr\left(\{Q^+,\bar\omega\phi\chi^+_{22}\}\right)-\raiz i\bar\omega
\tr\left(\phi F^+_{22}\right).
\eqn\ricardos
$$
Hence, 
$$
J(\bar\omega)=\{Q^+,\cdots\}+\underbrace{\int_X \bar\omega\tr\left(
\psi_{2\dalpha}\psi_2{}^{\dalpha}-\raiz i\phi F^+_{22}\right)}_{I(\bar\omega)}
,\qquad [Q^+,I(\bar\omega)]=0.
\eqn\canon
$$
Moreover, as the $m$ mass term does not enter in any of the above calculations,
the results also hold for $Q^+(m)$.

The preceding analysis implies that if we denote vacuum expectation values
in the twisted theory (which has topological symmetry $Q^{+}$and action $L$) 
by $\langle \ldots\rangle$, in the completely perturbed theory (with action 
$L+\Delta L(m,\omega)$ and symmetry $Q^{(m,\omega)}_1$) by 
$\langle \ldots\rangle_{m,\omega}$, and in the equivariantly extended theory 
(with action $L+\Delta L(m)$ and symmetry $Q^{+}(m)$) by  
$\langle \ldots\rangle_{m}$, the situation for the partition function is 
the following:
$$
\langle 1\rangle_{m,\omega}=
\left\langle \ex^{-J(\bar\omega)}\ex^{-\Delta L(m)}\right\rangle =
\left\langle \ex^{-J(\bar\omega)}\right\rangle_{m}. 
\eqn\merlin
$$
 
In the first equality we have discarded the $Q^{(m,\omega)}_1$-exact term 
\pinha . Notice that it is also possible, for the same reason, to discard the 
terms in \isa\ and \ennes . This leaves the $Q_1^{(m,\omega)}$-closed action
$L+\Delta^{(1)}+J(\bar\omega)$, where $\Delta^{(1)}$ are the mass terms
\sumathi\ and \fujita , i.e.
$$
\Delta^{(1)}= m\int_X\tr\left(\tilde\chi_{2\dalpha}\tilde\chi_1^\dalpha-
\raiz i\bar\phi[B^{+}_{12}+iC,B^{+}_{12}-iC]-2\raiz i
B^{+}_{11}[B^{+}_{22},\bar\phi]\right).
\eqn\petete
$$
Notice that $\Delta^{(1)}$ has ghost number $-2$, while $J(\bar\omega)$ has
ghost number $+2$. Also $L+\Delta^{(1)}$ is $Q^{+}(m)$-closed
(in fact it is $Q^{+}(m)$-exact up to a $\theta$-term). 
Hence, we can trade
$J(\bar\omega)$ for 
$\{Q^{+}(m),\cdots\}+I(\bar\omega)$ and discard the $Q^+(m)$-exact piece in
\ricardos . We are left with the action 
$$
\underbrace{L+\Delta^{(1)}}_{Q^+(m)-{\rm exact}}+\underbrace{I(\bar\omega)}_{
Q^+(m)-{\rm closed}}.
\eqn\menelao
$$
Now, as noted in [\wijmp] in a closely related context, $I(\bar\omega)$ (or
rather $J( \bar\omega)$) is the $F$-term of the chiral superfield $\Phi_3$; 
therefore, it cannot develop a vev if supersymmetry is to remain unbroken.
Strictly speaking, this applies to $\langle \psi_2\psi_2\rangle$. As for the
remaining term $\phi[B^{+}_{12}-iC,B^{+}_{22}]$, one can readily check that 
it vanishes on the moduli space. Hence,                     
$$
\left\langle \ex^{I(\bar\omega)}\right\rangle_m=\langle 1\rangle_m=\left\langle
\ex^{-\Delta^{(1)}}\right\rangle.
\eqn\unbelievable
$$

Finally, since $\Delta^{(1)}$ has ghost number $-2$, its vev in the original
theory must vanish as well, if the ghost number symmetry is to remain unbroken. 
Hence, under these assumptions, the partition function is invariant under the perturbation.


\vfill
\endpage


\chapter{Equivariant extension of the Thom form}

On K\"ahler manifolds there is a $U(1)$ symmetry acting on the moduli space. 
This symmetry was already noted in [\vafa] within the discussion of the 
vanishing theorem, which guarantees localization on the moduli space of ASD 
connections, but not  further use of it was made. 
We have discussed it in the previous section in connection with the mass
perturbations. Its action on the different fields is the following: 

$$
\cases{
B^{+}_{11}\to \ex^{-it}B^{+}_{11},\cr 
B^{+}_{12}-iC\to  \ex^{-it} (B^{+}_{12}-iC),\cr 
\tilde\psi^{+}_{11}\to  \ex^{-it}\tilde\psi^{+}_{11},\cr
\tilde\psi^{+}_{12}-{i\over2}\zeta
\to \ex^{-it}(\tilde\psi^{+}_{12}-{i\over2}\zeta),\cr 
\tilde\chi_{1\dalpha }\to \ex^{-it}\tilde\chi_{1\dalpha},\cr
\tilde H_{1\dalpha}\to \ex^{-it}\tilde H_{1\dalpha},\cr
}
\qquad
\cases{
B^{+}_{22}\to  \ex^{it}B^{+}_{22},\cr 
B^{+}_{12}+iC\to \ex^{it} (B^{+}_{12}+iC),\cr
\tilde\psi^{+}_{22}\to \ex^{it}\tilde\psi^{+}_{22},\cr
\tilde\psi^{+}_{12}+{i\over2}\zeta\to 
\ex^{it}(\tilde\psi^{+}_{12}+{i\over2}\zeta),\cr
\tilde\chi_{2\dalpha}\to \ex^{it}\tilde\chi_{2\dalpha},\cr
\tilde H_{2\dalpha}\to \ex^{it}\tilde H_{2\dalpha}.\cr
}
\eqn\circulo
$$ 

The gauge field $A$, the antighosts $\chi^{+}_{\alpha\beta}$ and $\eta$,
and the scalar fields $\phi$ and $\bar\phi$, carry no charge under this
$U(1)$. These transformations can be thought of as defining the
one-parameter flow associated to the action on the field space ${\cal M}$ of
the following vector field $X_{\cal M}\in T_{(A,B^+,C)}{\cal M}$:
$$
X_{\cal M}=\left(0,-iB^{+}_{11}, iB^{+}_{22}, -i(B^{+}_{12}-iC),
i(B^{+}_{12}+iC)\right).
\eqn\vectorcillo
$$ 

From the
viewpoint of the Mathai-Quillen formalism, the unperturbed twisted theory
provides a representation of the
${\cal G}$-equivariant de Rham cohomology (in the Cartan model) on the moduli
space. However, the formulation is not equivariant with respect to the $U(1)$
action. In other words, the perturbed action is not invariant (i.e. it is not
equivariantly closed) under the unperturbed twisted supercharge. On the other
hand, it is invariant under the perturbed twisted supercharge. In fact, the twisted supercharge $Q^{+}(m)$ of the perturbed
theory can be interpreted as the generator of the $
U(1)$-equivariant extension of the ${\cal G}$-equivariant de Rham cohomology on
the moduli space. This connection between  massive extensions of twisted 
supersymmetric
theories and equivariant cohomology was pointed out in 
\REF\koreas{S. Hyun, J. Park and J.-S. Park, ``$N=2$ Supersymmetric QCD and 
Four-Manifolds; (I) the Donaldson and the Seiberg-Witten Invariants", 
hep-th/9508162.}
[\koreas], in the context of the non-Abelian monopole theory with massive
hypermultiplets; it was subsequently exploited in 
\REF\zzeta{J. M. F. Labastida and M. Mari\~no, ``Twisted $N=2$ Supersymmetry with Central Charge and Equivariant Cohomology"\journal\cmp&185 (97)37; hep-th/9603169.}
[\zzeta], where the explicit construction leading to the idea of the 
equivariant extension was carried out in detail. In what follows, we will try 
to adapt the construction in [\zzeta] to our
problem. We intend to be as sketchy as possible, and therefore refer the reader to the work cited above for the minute details of the construction. 

The idea underlying the construction is the following. Prior to the perturbation, we have a topological field theory which admits a Mathai-Quillen description
with BRST charge $Q^{+}$. This means, among other things, that the
corresponding Lagrangian is a $Q^{+}$-commutator.  After adding the mass
terms proportional to $m$, it is possible to modify the
$Q^{+}$ transformation laws so that the perturbed Lagrangian can be written 
as a $Q^{+}(m)$-commutator as well, where $Q^{+}(m)$ are the modified
topological transformations. In view of this, it would be tempting to assume 
that there has to be a standard Mathai-Quillen construction associated to the
new topological theory. However, the perturbation has not changed the
geometrical setting of the problem, so there is a priori no reason why the 
Mathai-Quillen formulation should change at all. In fact, it does not,
and it turns out that the perturbed theory admits no standard
Mathai-Quillen formulation. However, as pointed out in [\zzeta], the 
formalism allows a natural generalization in those situations 
in which there is an additional symmetry group acting on the moduli space. The
geometrical construction involved is an equivariant extension of the Thom form
of ${\cal E}$ within the framework of the Mathai-Quillen formalism.    
 
The Mathai-Quillen formalism provides an explicit representative of the Thom
form of the oriented vector bundle ${\cal E}={\cal M}\times_{\cal G}{\cal F}$. 
The bundle ${\cal E}$ is awkward to work with, and it is preferable to work
equivariantly, i.e. to regard ${\cal E}$ explicitly as an associated vector 
bundle to the ${\cal G}$-principal vector bundle ${\cal M}\times {\cal F}\to
{\cal E}$. The Mathai-Quillen representative of the Thom form of ${\cal E}$ is
${\cal G}$-equivariantly closed and basic on ${\cal M}\times {\cal F}$ (and
hence descends naturally to ${\cal E}$). In the Weil model for the ${\cal
G}$-equivariant cohomology of ${\cal E}$, the Mathai-Quillen form is an
element in ${\cal W}({\bf g})\otimes\Omega^{*}(F)$ (${\cal W}({\bf g})$ 
is the Weil algebra of $G$) given by [\moore]:
$$
U=\ex^{-\vert x\vert^2}\int D\rho\,{\hbox{\rm exp}}\Biggl ({1\over4}\rho_i 
K_{ij}\rho_j +i\rho_i(dx_i+\theta_{ij}x_j)\Biggr).
\eqn\thom
$$
In \thom\ $x_i$ are orthonormal coordinates on the fibre ${\cal F}$, and 
$dx_i$ are their corresponding differentials. The $\rho_i$ are Grassmann
orthonormal coordinates for the fibre, while $K$ and $\theta$ are the
generators of ${\cal W}({\bf g})$. The Chern-Weil homomorphism, which
essentially substitutes the universal realizations $K$ and $\theta$ by the 
actual curvature and connection in ${\cal M}\times {\cal F}$, gives the link
between the Universal representative $U$ and the Thom form $\Phi({\cal E})$.  
The important point is that while $U$ is ${\cal G}$-equivariantly closed by 
construction, it is not equivariantly closed with respect to the $U(1)$
action. It seems natural to look for a redefinition of the representative
\thom , which is  $U(1)$-equivariantly closed. The equivariant
extension of $U$ with respect to the $U(1)$ action simply amounts to finding a
suitable form $p$ such that $U+p$ is  $U(1)$-equivariantly closed. Within the framework of the Mathai-Quillen formalism
this amounts to replacing the curvature $K$ with a new equivariant curvature 
$K_{U(1)}$ [\zzeta], which is just the original curvature $2$-form $K$ plus
an operator $L_\Lambda$ involving the infinitesimal $U(1)$ action and the
connection $1$-form $\theta$. In the Cartan model, which is the best suited
to topological field theories, the connection form is set to zero, and hence
the equivariant extension of the curvature is just the original one plus an
operator implementing the infinitesimal $U(1)$ action. This may sound rather 
abstract, so we now proceed to the
actual construction. The main ingredients are a $U(1)$ action defined on the
moduli space and the fibre ${\cal F}$, under which the metrics on both the
moduli space and the fibre must be invariant, while the section $s:{\cal M}\to
{\cal V}$ has to transform equivariantly; that is, if $\phi^{\cal M}_t$ and 
$\phi^{\cal F}_t$ denote the action of $U(1)$ on ${\cal M}$ and 
${\cal F}$ respectively, then 
$$
s\cdot\phi^{\cal M}_t=\phi^{\cal F}_t\cdot s 
\eqn\hola
$$
This can be easily verified in the present problem in view of the form of $s$
\ecuaciones\ and the $U(1)$ actions \circulo . As for the metrics, it suffices
to show that for two vector fields $(0,X^{+},x)$ and $(0,Y^{+},y)$, their
scalar product is invariant under the $U(1)$ action \circulo . According to our
conventions [\noso], $\tr X^{+}_{\alpha\beta}Y^{+\alpha\beta}=-4\tr X\wedge
*Y$, so a natural definition for the metric on the field space would be as
follows ($\langle | \rangle$ denotes the scalar product on $T{\cal M}$):
$$
\eqalign{
&\left\langle (0,X^{+},x)|(0,Y^{+},y)\right\rangle = -\int_X \tr \left( X^{+}_{\alpha\beta}
Y^{+\alpha\beta}\right)+2\int_X \tr *(xy)\cr &= -\int_X\tr\,\big(X^{+}_{11}
Y^{+}_{22}+X^{+}_{22}Y^{+}_{11}\big)+\int_X \tr\, \big[ (X^{+}_{12}+ix)
(Y^{+}_{12}-iy)+(X^{+}_{12}-ix)(Y^{+}_{12}+iy)\big],
\cr}
\eqn\cernch
$$
which is indeed invariant under the $U(1)$ action. 

To incorporate the $U(1)$
action to the Mathai-Quillen construction sketched in section 2.2, we
modify the $Q^{+}$ transformations of the ghosts and the
auxiliary fields charged under 
$U(1)$ by replacing the curvature $\phi$ with its equivariant extension 
$\phi(t)=\phi+{\cal L}_t$, where  ${\cal L}_t$ generates on the fields an 
infinitesimal $U(1)$ transformation. 
According to \circulo , this affects only   
$\tilde\psi^{+}_{\alpha\beta}$, $\zeta$ and $\tilde H_{\alpha\dalpha}$. In
view of \papagayo , the new transformations read:
$$ 
\eqalign{
\{Q^{+}(t),\tilde\psi^{+}_{11}\}&=2i([ B^{+}_{11},\phi]-itB^{+}_{11}),\cr
\{Q^{+}(t), \tilde\psi^{+}_{22}\}&=2i([B^{+}_{22},\phi]+ itB^{+}_{22}),\cr
\left\{Q^{+}(t), \tilde\psi^{+}_{12}\pm {i\over2}\zeta\right\}&=
2i\bigl([B^{+}_{12}\pm iC,\phi]\pm it(B^{+}_{12}\pm iC)\bigr),\cr
[Q^{+}(t),\tilde H_{1\dalpha}]&=
2\raiz i([\tilde\chi_{1\dalpha},\phi]-it\tilde\chi_{1\dalpha})-\raiz 
[Q^{+},s_{1\dalpha}]
,\cr
[Q^{+}(t),\tilde H_{2\dalpha}]&=
2\raiz i([\tilde\chi_{2\dalpha},\phi]+it\tilde\chi_{2\dalpha})-\raiz 
[Q^{+},s_{2\dalpha}],\cr
}
\eqn\final
$$
If we set $t=-{m\over\raiz}$ we see that eqs. \final\ reduce precisely to 
the $Q^{+}(m)$ transformations \silla\ and \papaya . The transformations
\final , when applied to the gauge fermion \Mazinger , reproduce the original
unperturbed action plus the mass terms \sumathi\ and \fujita . To
reproduce the remaining mass terms we note that, as is standard in
topological (cohomological) field theories, there remains
the possibility of adding to the action a $Q^{+}(t)$-exact piece without
--hopefully-- disturbing the theory. As discussed in [\zzeta], the requisite piece 
can be interpreted as the equivariantly-exact differential form which is conventionally added to prove localization in equivariant integration. It has the form $\{Q^{+}(t),\omega_{X_{\cal M}}\}$, where $\omega_{X_{\cal M}}$ is the differential form given by $\omega_{X_{\cal M}}(Y)=\langle X_{\cal M}|Y\rangle$, $Y$ being a vector field on ${\cal M}$. In view of the form of the vector field $X_{\cal M}$ \vectorcillo\ and of the metric \cernch , and keeping in mind that the ghosts $(\psi, \tilde\psi^{+},\zeta)$ provide a basis of differential forms on ${\cal M}$, this form gives a contribution 
$$
\eqalign{
\Biggl\{ Q^{+}(t), -{it\over2}\int_X\tr\Bigr(\tilde\psi^{+}_{22}(-iB^{+}_{11})+\tilde\psi
^{+}_{11}(iB^{+}_{22})\Bigl)+&{it\over2}\int_X\tr\bigg((-i)(B^{+}_{12}-iC)
\left(\tilde\psi^{+}_{12}+{i\over2}\zeta\right)\cr &+i(B^{+}_{12}+iC)
\left(\tilde\psi^{+}_{12}-{i\over2}\zeta\right)\bigg)\Biggr\}.
\cr}
\eqn\morel
$$

But these are precisely the terms
\Rimbaud\ and
\Apollinaire , which as we have seen give correctly the remaining mass terms. 

\vfill\eject


\chapter{Conclusions}

The analysis presented in this paper supports the assumption made in [\vafa] in the context of the abstract approach applied to one of the twisted $N=4$ supersymmetric gauge theories. Namely, we have shown that on K\"ahler four-manifolds the partition function of the Vafa-Witten theory, the only observable leading to topological invariants, remains unchanged under a mass perturbation. This result depends crucially upon the fact that the ghost number symmetry of the theory is non-anomalous. Likewise, we have shown that the mass-perturbed theory (with $\omega=0$) can be regarded as the equivariant extension of the original theory with respect to the $U(1)$ action on the moduli space described in [\vafa].

As was stated in the introduction, the use of the abstract approach in the context of topological quantum field theory is very interesting, because it relies entirely on the properties of physical $N=1$ supersymmetric gauge theories. Thus, it constitutes an important (and truly independent) test of these properties. The predictions in this approach can be tested by confronting them to known mathematical results, or to alternative results obtained in the concrete approach.  
The results in the framework of the concrete approach recently presented in [\wimoore] constitute a very fruitful arena to test predictions based on the abstract approach. In this sense, a wide context is now available to test the properties that are usually attributed to physical $N=1$ supersymmetric gauge theories. To date, only three models have been studied within the abstract approach: the Donaldson-Witten theory with gauge group $SU(2)$, 
the non-Abelian monopole theory with gauge group $SU(2)$ and a 
matter multiplet in the fundamental representation, and the Vafa-Witten theory. Other models, as for example those involving higher-rank groups and/or an extended set of hypermultiplets, should also be considered. Our analysis shows that the validity of the abstract approach has to be analysed case by case.

The second twist of the $N=4$ supersymmetric $SU(2)$ gauge theory seems to be  quite a promising example. This theory has an anomaly in the ghost number symmetry equal to $-{3\over4}(2\chi + 3\sigma)$ for $SU(2)$ [\noso]. On K\"ahler manifolds, this is proportional to the square of the canonical class ($K\cdot K= 2\chi+3\sigma$) and therefore vanishes on hyper-K\"ahler manifolds. This is as it should be, for the physical and the twisted theory coincide on hyper-K\"ahler manifolds [\wijmp], and therefore the second twist should be equivalent to the Vafa-Witten theory, which is anomaly-free. On more general four-manifolds, this is no longer the case, and in order to compute non-trivial topological correlators one has to insert operators whose overall ghost number matches the anomaly of the theory. Notice that since the anomaly does not depend on the instanton number, there is only a finite number (if any) of non-vanishing correlation functions. One could in principle try to compute   
these topological observables in the pure (i.e. massless) twisted theory. As there is no equivalent of the $u$-plane description for the low-energy dynamics of $N=4$ theories, the concrete approach does not apply to this case. As for the abstract approach, the $N=1$ low-energy theory which corresponds to $N=4$ perturbed with a mass term for one of the chiral superfields, has a continuum of vacuum states in different phases, and therefore it is not very useful for making explicit computations.

For the mass-deformed twisted theory ($N=4$ with masses for two of the chiral superfields) the situation is certainly different. The corresponding physical theory has $N=2$ supersymmetry and its low-energy behaviour is known [\swotro]. There is a definite picture of the structure of singularities and of the symmetries governing the dynamics on the $u$-plane, and it is therefore possible to make explicit computations within the concrete approach. As for the twisted theory, unlike the Vafa-Witten theory, the mass perturbation makes sense on any arbitrary spin four-manifold. The perturbation preserves the unique topological symmetry of the theory, and in fact it can be shown, by extending the construction presented in [\zzeta] for the theory of non-Abelian monopoles, that the structure of the perturbation is dictated by an equivariant extension with respect to a $U(1)$ action which is a symmetry of the non-Abelian adjoint monopole equations. However, as the ghost number symmetry is generally anomalous, one should expect the correlation functions to depend non-trivially on the mass parameter $m$. Of course, on hyper-K\"ahler manifolds one should recover the results of [\vafa]. In particular, the generating function for the topological correlators should converge to the partition function presented there. As regards the abstract approach, the vacuum structure of the $N=1$ effective theory is known (and we have discussed it above): there are three isolated vacua with a definite pattern of symmetries relating them. The space-time-dependent mass term, which breaks $N=2$ down to $N=1$, cannot simply be dropped as in the Vafa-Witten theory. Rather, as this term is essentially one of the observables of the theory, the effect of the perturbation can be absorbed, as in [\wijmp] or [\marpol], in a redefinition of the parameters in the generating function. We expect to address these and other related issues in future work.

\vskip2cm


\ack
We are specially indebted to M. Mari\~no for enlightening discussions 
and many useful remarks. This work was supported in part by DGICYT under grant PB93-0344, and by the EU Commission  under TMR grant FMAX-CT96-0012.

\vfill
\endpage
\refout
\end

(Esta parte ha sido censurada por la violencia explicita de los argumentos que contiene)            
               
-----------------------------------------------------------------
LOCALIZATION

The perturbed theory (with $\omega=0$) is expected to localize on the 
fixed points of the $Q^{+}(m)$ symmetry. In view of \sandia ,\pavia, \sandy\ 
and \silla , the $Q^{+}(m)$ fixed points yield the equations:
$$
\cases{
i[ B^{+}_{11},\phi]-{1\over\raiz} m B^{+}_{11}=0,\cr
i[ B^{+}_{22},\phi]+{1\over\raiz} m B^{+}_{22}=0,\cr
i[B^{+}_{12}\pm iC,\phi]\pm{1\over\raiz} m (B^{+}_{12}\pm iC)=0,\cr
\deriv_{\alpha\dalpha}\phi=0= [\phi,\bar\phi],\cr}
\qquad
\cases{
F^{+}_{11}-2[B^{+}_{12}+iC,B^{+}_{11}]=0,\cr
F^{+}_{22}+2[B^{+}_{12}-iC,B^{+}_{22}]=0,\cr
F^{+}_{12}+[B^{+}_{12}-iC,B^{+}_{12}+iC]+[B^{+}_{11}, B^{+}_{22}]=0,\cr
D_{1\dalpha} B^{+}_{22}-D_{2\dalpha}(B^{+}_{12}+iC)=0,\cr
D_{2\dalpha} B^{+}_{11}-D_{1\dalpha}(B^{+}_{12}-iC)=0,\cr
} 
\eqn\wagner
$$
Notice that the set of equations on the right is exactly the same 
as before \masecuaciones . Hence, the general arguments leading to 
the vanishing theorem are not affected at all by the
perturbation. The moduli space defined by \wagner\ has 
two branches. On the one hand, the equations \wagner\ admit the trivial 
solution $B^{+}_{\alpha\beta}=0=C$, which leads to 
$F^{+}_{\alpha\beta}=0$ and $\deriv\phi=0=[\phi,\bar\phi]$. But this 
is precisely the moduli space  of ASD connections which
arises in Donaldson-Witten theory [\tqft]. On the other hand, in presence 
of the $m$ mass terms 
the moduli space develops a second branch in which $\phi\not=0$ (that is, 
the gauge connection is reducible) and $B^{+}_{11}$, $B^{+}_{22}$
and $B^{+}_{12}\pm iC$ don't vanish either. 
     
                   ---------------------------